1 **Variety of stylolites morphologies and statistical characterization of the**

2 **amount of heterogeneities in the rock**

5 **Alexandre Brouste**

6 Laboratoire de Modélisation et de Calcul, Université Joseph Fourier, BP 53, 38041 Grenoble, France

7 **François Renard**[*]

8 Laboratoire de Géophysique Interne et Tectonophysique, CNRS, Université Joseph Fourier, BP 53, 38041

9 Grenoble, France & Physics of Geological Processes, University of Oslo, Norway

10 **Jean-Pierre Gratier**

11 Laboratoire de Géophysique Interne et Tectonophysique, Université Joseph Fourier, BP 53, 38041 Grenoble,

12 France

13 **Jean Schmittbuhl**

14 UMR 7516, Institut de Physique du Globe de Strasbourg, 5 rue René Descartes, F-67084 Strasbourg cedex,

15 France

17 [*]Corresponding author:

18 francois.renard@ujf-grenoble.fr, Phone +33 476 82 80 88, Fax; +33 476 82 81 01

**Abstract.** The surface roughness of several stylolites in limestones was measured using high resolution laser profilometry. The 1D signals obtained were statistically analyzed to determine the scaling behavior and calculate a roughness exponent, also called Hurst exponent. Statistical methods based on the characterization of a single Hurst exponent imply strong assumptions on the mathematical characteristics of the signal: the derivative of the signal (or local increments) should be stationary and have finite variance. The analysis of the measured stylolites show that these properties are not always verified simultaneously. The stylolite profiles show persistence and jumps and several stylolites are not regular, with alternating regular and irregular portions. A new statistical method is proposed here, based on a non-stationary but Gaussian model, to estimate the roughness of the profiles and quantify the heterogeneity of stylolites. This statistical method is based on two parameters: the local roughness ($H$) which describes the local amplitude of the stylolite, and the amount of irregularities on the signal ($\mu$), which can be linked to the heterogeneities initially present in the rock before the stylolite formed. Using this technique, a classification of the stylolites in two families is proposed: those for which the morphology is homogeneous everywhere and those with alternating regular and irregular portions.

*Key words*: stylolites, roughness, scaling analysis, heterogeneity

## 1. Introduction

37

38 The geometrical characterization of rough profiles or surfaces is a widespread problem in

39 various geological examples such as erosion patterns (Dunne, 1980; Cerasi et al., 1995),

40 multiphase fluid percolation in porous rocks (Rubio et al., 1989), fractures (Schmittbuhl et

41 al., 1993), or stylolites (Renard et al., 2004). In these studies, the scaling behavior of

42 various data sets was investigated, showing that the statistics at one scale could be

43 extrapolated to another scale using a power law relationship.

44 For a self-affine function $h(x)$, a scaling relationship is defined when the signal follows a

45 power law relationship under a dilation of a factor $\lambda$

46
$$h(\lambda x) = \lambda^D h(x) \qquad (1)$$

47 where $x$ is the spatial coordinate and $h$ is a scalar field, $\lambda$ is the scaling scalar, and $D$ is the

48 scaling exponent.

49 Applying this property to 1D discrete signals, involves working on the increments $\delta h(x)$ of

50 the function $h$. The self-similar property of a 1D data set $h(x)$ emerges when the increments

51 of the signal follows

52
$$\delta(h(\lambda x)) = \lambda^H \delta(h(x)) \qquad (2)$$

53 where $H$ is the so-called Hurst exponent (Feder, 1988; Meakin, 1998).

54 This scaling approach is based on two assumptions on the mathematical properties of the

55 signal. First, increments of the signal have finite variance distribution and, second they are

56 stationary, which means that the statistics are independent of the position along the signal.

57 In the case of a signal with increments that follow a Gaussian distribution (that has a finite

58 variance by definition), the roughness of the signal can be deduced from the scaling

59 exponent.

60 For a function $h(x)$ with the property:

61 $$|h(x) - h(y)| \leq C|x - y|^{H_0} \, ,$$ (3)

62 where $x$ and $y$ are two different points along the signal and $C$ is a constant, $H_0$ is defined as

63 the Hölder exponent (Daubechies, 1992). When the increments of a signal are Gaussian and

64 stationary, the Hölder exponent is equal to the Hurst exponent.

65 In this contribution, the assumption of Gaussian stationary increments of several 1D data

66 sets is tested, based on roughness measurements of various stylolites in limestones. We

67 show that these profiles do not verify the Gaussian stationary increments property, and we

68 propose a new technique to characterize the statistics of these signals by introducing two

69 parameters: the *localized* roughness exponent $H$, and a second parameter $\mu$, which

70 characterizes the quantity of irregularities in the system at all scales. Applied to stylolites,

71 this parameter can be used to quantify the degree of heterogeneity in the rock initially

72 present before the stylolitization process. We also show that heterogeneities have an effect

73 only above a millimeter scale.

74 We first present some examples showing how heterogeneities determine the location of

75 some stylolite peaks. Then the two-parameter statistical description of stylolite roughness is

76 used to help characterize such heterogeneities.

77 **2. The roughness of stylolites**

78 *2.1. Self-similar scaling of stylolites*

79 Stylolites are rough surfaces that develop by stress-enhanced dissolution in crustal rocks

80 (Dunnington, 1954; Park and Schot, 1968; Bathurst, 1971; Bayly, 1986). Anticrack models

81 have been proposed to describe their initial stage of nucleation and propagation as a flat

82 interface (Fletcher and Pollard, 1981; Koehn et al., 2003; Katsman and Aharonov, 2006).

83 With time, the stylolites roughen and acquire their typical wavy geometry (Figures 1, 2).

84    The wide range of morphological geometries of such surfaces makes them difficult to

85    characterize using a simple scaling approach. However, it has been shown that stylolites

86    have self-similar scaling properties (Karcz and Scholz, 2002; Renard et al., 2004;

87    Schmittbuhl et al., 2004). These studies are based on the assumption that the morphological

88    statistics of the stylolites do not vary laterally along the plane of the interface.

89    Here, the topography of stylolites in limestones was measured using high-resolution laser

90    profilometers that acquire (1+1)D roughness profiles (Figure 2). Some stylolites were split

91    open to reveal the complex 2D geometry of their surface. Using this method, described in

92    Renard et al., (2004), (2+1)D maps of stylolite roughness can be obtained with an accuracy

93    of up to 0.003 mm, on a regular grid of 0.03 to 0.125 mm depending on the kind of

94    profilometer used. The (2+1)D maps were built by combining (1+1)D profiles on a square

95    grid with a constant discretization interval. For each stylolite surface, the result is a (2+1)D

96    height field from which the mean plane was removed by a least-square method.

97    Using these data, stylolitic 1D profiles were found to show two different self-affine regimes

98    at large and small length scales (Figure 3). Two signal processing techniques were used: the

99    Fourier Power Spectrum (FPS) and the Averaged Wavelet Coefficient (AWC).

100   FPS decomposition techniques are standard tools used to characterize the scaling behaviour

101   of stationary increments signals (Kahane and Lamarié-Rieusset, 1998). Assuming finite

102   variance stationary increments of a signal, the Hurst exponent $H$ (eq. 2) can be deduced

103   from the power-law behaviour of the Fourier Power Spectrum with

104   $$FPS(k) \propto k^{-1-2H} \qquad\qquad (4)$$

105   where $k$ is the wave number, the inverse of the wavelength (Barabási and Stanley, 1995).

106   Wavelet series (or wavelet decompositions) constitute a powerful tool for processing

107   signals in which different scales are combined (Meyer and Roques, 1993). Various signals

108   can be reconstructed knowing the coefficients of their wavelet decomposition, and for

109 compactly supported wavelets (Daubechies, 1992) any 1D profile, *h(x)*, can be decomposed

110 into a wavelet series having the following summation:

111
$$h(x) = \sum_{j=0}^{+\infty} \sum_{i=0}^{2^j - 1} c_{i,j} \psi\left(2^j x - i\right) \qquad (5)$$

112 where $c_{j,i}$ are the wavelet coefficients indexed by (*j*,*i*) and $\psi$ is the so-called mother

113 wavelet (generating all the wavelets by expansion of a factor $2^j$ and by a translation *i*).

114 Using this method, the self-similar behaviour of a signal emerges as the average wavelet

115 coefficient AWC satisfies:

116
$$AWC\ (l) \propto l^{H + 0.5}, \qquad (6)$$

117 where *l* is the spatial wavelength (Simonsen *et al.*, 1998).

118 These two techniques provide a scaling relationship and the Hurst exponent is directly

119 related to the slope of the spectra. In the case of a signal with Gaussian and stationary

120 increments, the Hölder exponent is equal to the Hurst exponent.

121 In stylolites, these two signal processing techniques give the same Hurst exponent (eq. 2),

122 *H = 0.5* for the large length scales and *H = 1.1* for small length scales (Figure 3, see also

123 Renard *et al.*, 2004).

124 The measurements also show that a sharp cross-over length scale close to the millimeter

125 scale separates the two regimes. This characteristic length scale has been interpreted as a

126 crossover length emerging from the competition between two forces: surface tension

127 dominates at small wavelengths, whereas elastic interactions dominate at large wavelengths

128 (Renard et al., 2004; Schmittbuhl et al., 2004).

129 Using the same data sets, it can also be shown that a stylolite can be wavy at one point and

130 rather flat at another point (Figure 2), suggesting that the statistical properties vary along

131 the profiles. Therefore, the Gaussian stationary increments hypothesis must be called into

132    question. This spatial variation in statistical properties along a single stylolite is not
133    accounted for in current models of stylolite roughening.

134    *2.2. Heterogeneities along stylolites*

135    Various examples both from nature and experiments show that heterogeneities in rocks help
136    either to localize dissolution pits or to deflect the dissolution surface along a single stylolite
137    at all scales. Figure 4a shows experimental microstylolites along quartz grains (Gratier et
138    al., 2005). Dissolution pits (Figure 4b) are systematically located at the bottom of each
139    conical-shaped stylolite structure. Due to the fit of the two opposite grain surfaces, the pits
140    of the lower grain stylolite surface are located just in front of the stylolitic peak of the upper
141    grain and vice versa. The explanation is that pits develop at intersections of crystal
142    dislocations with the grain surface and determine the stylolite peak location.

143    Figure 4c shows the indenting of a mineral (quartz) by another mineral (mica). In this case,
144    the mica grains along the dissolution surface are responsible for the local dissolution peaks.
145    Mica distribution determines the location of the peaks location.

146    The same geometry may be observed along columnar stylolites in limestones (Figure 4d).
147    However, the interpretation is different as the two parts of the rock have the same
148    composition. In this case, the geometry of the columnar stylolite is probably determined by
149    preexisting micro-fractures as is clearly the case in the example shown in Figure 4e where a
150    fracture controls the shape of the peak. Finally, Figure 4f shows several dissolution seams
151    that are deflected by hard objects: pyrite (black) or quartz pressure shadows (white). In this
152    case, the hard objects located in the dissolution plane deflect it, thereby contributing to
153    roughening of the dissolution surface.

154    All these examples show that the location of some stylolite peaks is not purely random but
155    rather partially controlled by the distribution of heterogeneities. The statistical properties of
156    stylolites should depend on the distribution of these heterogeneities, and therefore vary in

157 space along a single stylolite. It would appear relevant to integrate the presence of non-
158 uniformly distributed heterogeneities at all scales in the modeling of stylolites and test their
159 potential effect on the final geometry.

## 3. A two-parameter statistical description of the roughness of 1D stylolite
161 profiles

162 The wide range of morphologies of stylolites (Figure 1) and the alternating smooth and
163 irregular portions of the same stylolite (Figures 2, 5a), suggests that the Gaussian stationary
164 increments assumption should be tested. In this section we show that it is not possible to
165 obtain all stylolite morphologies from a single parameter scaling relationship (e. g. a Hurst
166 exponent).

167 Figure 5b represents the increments of a 1D stylolite. These increments are calculated as the
168 height difference between two successive points, and therefore represent a first order
169 derivative of the original signal of Figure 5a. In this incremental signal, the existence of
170 many large jumps and long tails in the histogram (Figure 5c) differentiate the signal from a
171 synthetic fractional Gaussian noise signal (Figure 5h). Therefore, the Gaussian self-similar
172 stationary increments property can be excluded for stylolite signals and a simple scaling
173 relationship using a single Hurst exponent is not sufficient to explain the measured signals.
174 The following section proposes a new technique that can accommodate the large jumps of
175 Figure 5b so that it can be applied to stylolites. This analysis has been tested on all the
176 available stylolites surfaces, and show similar properties.

177 *3.1. The Simple Branching Process Wavelet Series method*

178 Mathematicians commonly use two different techniques to deal with the large jumps
179 similar to those shown in Figure 5b. The first technique is to select a non-Gaussian self-
180 similar stationary increment model with infinite variance, also called stable Lévy motion

181 (Samorodnitsky and Taqqu, 1994). Stable Lévy motions contain two parameters: the
182 frequency of the jumps and the average size of these jumps. Applied to stylolites,
183 microfractures densities in the rocks can be associated with the frequency of jumps for
184 instance and estimated by specific methods. However, in such models, the roughness
185 cannot be identified from the scaling relationship because the roughness and the scaling
186 exponents are not similar. The Lévy models are avoided in the following discussion.

187 The second technique is a non-stationnary Gaussian model with scaling properties, where
188 the roughness can be estimated. According to Samorodnitsky and Taqqu (1994), neither of
189 these techniques is superior to the other.. In the following section, the non-stationary
190 Gaussian model is used and referred to as the Simple Branching Process Wavelet Series (in
191 short SBPWS).

192 *3.2. Construction of SBPWS profiles in one dimension*

193 Simple Branching Processes (also called Galton-Watson processes, see Harris, 1969) are
194 stochastic trees built by an incremental branching process at all scales. In the case of Simple
195 Branching Process Wavelet Series (SBPWS), each node of the tree has the same probability
196 of having either one or two branches (see Figure 6a). In the following, $1 <$
197 $\mu < 2$ corresponds to the average number of sons at each node. For a node of the tree, (2-
198 $\mu$) represents the probability of having only one branch.

199 SBPWS models are particular random *lacunary* wavelet series (Jaffard, 2000) based on
200 simple branching processes. Lacunary refers to the property that only a small number of
201 coefficients in the series are non-vanishing, more precisely those indexed by an elementary
202 branching process and corresponding to the branches of Figure 6a. SBPWS is defined by:

203
$$SBPWS(x) = \sum_{j=0}^{\infty} 2^{-jH} \sum_{i \in \Lambda(\mu)} \varepsilon_{j,i} \, \psi(2^j x - i) \qquad (7)$$

where $x$ is the spatial coordinate, $H$ is the fractional parameter, $\Lambda(\mu)$ is the elementary branching process of parameter $\mu$, $\varepsilon_{j,i}$ are a family of independent Gaussian standard random variables and $\psi$ is a wavelet-like function.

Only wavelets with coefficients indexed by the stochastic sub-tree $\Lambda$ (of non-vanishing coefficients) contribute to the roughening of the initial flat profile (see Figure 6b). Therefore, the stochastic tree process $\Lambda$ locally deforms the 1D profile, at all the tree branches.

In this model, elementary forms of the deformation are given by the shape of the mother-function $\psi$. A difficulty with modeling a stylolitic structure is to choose the function $\psi$, which corresponds to the shape of each dissolution increment. However, it has been shown that the statistics of a simulated signal do not depend on the shape of $\psi$, as long as this function has the same property as an individual wavelet (Brouste, 2006).

In nature, the stylolite shape varies from columnar to conical (Figures 1, 4) and these two kinds of shape might be related to the shape of microscopic increment of dissolution: either rectangular for columnar stylolites, or triangular for the conical ones. As a consequence, a choice must be made in the mathematical modeling between rectangular or triangular increments or a specific parameter used that may express all the intermediary shapes. Moreover, columnar stylolites are rather specific, being associated either with microfractures (Figure 4d and 4e) or with non-consolidated material (Gratier et al., 2005). In order to avoid the use of a third parameter, the shape of the function $\psi$, which might hide the effect of the two other parameters, a triangular function was chosen for $\psi$ (Figure 6b, inset). Note that the choice of the shape of this function $\psi$ does not modify the statistical properties of the synthetic signal.

227   The natural stylolites that were examined in this study can be modeled with such an

228   elementary triangular shape. By varying the parameters $H$ and $\mu$, one can generate synthetic

229   profiles that have stylolite-like patterns (Figure 7; Appendix A gives the algorithm to build

230   these synthetic stylolites). These synthetic profiles, unlike those generated by previous

231   models, exhibit two important properties of the natural stylolites:

232   i) the variability of the roughness between independent stylolite profiles;

233   ii) the variability of the roughness within a single profile, with alternating regular and

234   irregular portions.

235   *3.3. Parameters H and $\mu$*

236   The parameters $H$ and $\mu$ have distinct visual effects on the synthetic profiles. The

237   irregularities on the whole profile are quantified by the parameter $\mu$: for instance, at the n[th]

238   order branches, there are, on average, $\mu^n$ non-vanishing coefficients and then $\mu^n$ branches

239   of the tree, corresponding to $\mu^n$ stages of deformation of the initially flat profile. When

240   $\mu$ is close to 2, there are irregularities everywhere along the profile. When $\mu$ decreases to 1,

241   there are alternating irregular and regular portions along the profile. Finally, when $\mu$ is

242   equal to 1, there are no more irregularities along the signal.

243   The amplitude of the deformation (only where it is deformed) depends on the scale, on a

244   random Gaussian variable, and on a fractional exponent $H$ that can be considered to be a

245   local roughness parameter. In this sense, $H$ is indicative of the nature of the irregularity and

246   the amplitude of the profile variations. When $H$ tends to 0 the profile is irregular and looks

247   "noisy". This property is also called antipersistence: locally a valley in the signal has a

248   greater probability of being followed by a hill. When $H$ is close to 1, the profile roughness

249   is smoother and a valley or a hill in the signal tends to extend locally. This property is

250   called persistence (Meakin, 1998).

251   *3.4. Measurements of H and μ on a 1D data set*

252   As stated previously, the SBPWS have scaling properties that no longer involve a unique

253   stationary Hurst exponent. SBPWS provides self-affine behavior either in the 1D Average

254   Wavelet Coefficient technique or in the Fourier Power Spectrum, and is defined by a

255   power-law in both scale and frequency domains, respectively (Brouste, 2006):

256   $$AWC\,(\,l\,) \propto l^{\,1 - \log_2 \mu\,/\,2 + H}$$   (8)

257   and

258   $$FPS\,(\,k\,) \propto k^{\,-2 + \log_2 \mu - 2H}$$   (9)

259   where $H$ and $\mu$ are the two parameters of the SBPWS method. When $\mu = 2$, equations (8)

260   and (9) are reduced to the Gaussian stationary case described in equations (4) and (6).

261   Note that in the SBPWS method, the values of $H$ and $\mu$ cannot be determined by a simple

262   regression to the 1D Fourier and AWC spectra, as done previously by Renard et al. (2004),

263   because the following system of equations, whose determinant is equal to zero, is

264   underdetermined:

265   $$\begin{cases} 2H - 2 + \log_2 \mu = a \\ H - 1 + \log_2 \mu\,/\,2 = b \end{cases}$$   (10)

266   Here $a$ and $b$ are the slopes measured by linear regression on the FPS spectrum and on the

267   AWC spectrum, respectively.

268   Therefore, a more complex tool must be used, such as the s-generalized variations method

269   (Istas and Lang, 1997) to obtain estimated values of $H$ and $\mu$ at large and small length

270   scales. This method, detailed in Appendix B, was applied to estimate $H$ and $\mu$ in the

271   stylolites that were measured (Table 1).

272   **4. Application to natural stylolites**

273   *4.1. Parameters H and μ for the stylolites*

274    To estimate the parameters $\mu$ and $H$, from both sides of the cross-over length scale, it is

275    necessary to observe how the estimators of Appendix B behave when the length scale

276    decreases (as $n$ increases), from large scales to small scales through the cross-over length

277    scale (Figure 8a-b). Large length scales values are taken at the cross-over length scale and

278    small ones are taken at the discretization scale in order to use the greatest number of points

279    in the two different patterns.

280    The results presented in Table 1 are based on averaged estimations of a series of 256 to 512

281    parallel stylolite profiles, each profile being regularly discretized on 512 to 1024 points.

282    This gives the large length scale and the small length scale parameters $H$ and $\mu$ for all the

283    stylolites that have been measured.

284    *4.2. Geometrical characterization*

285    Most of the information on $\mu$ and $H$ variability belongs to the large length scale parameters

286    (see Table 1). In fact, small length scale parameters have almost similar values ($\mu$ from 1.2

287    to 1.4 and $H$ from 0.6 to 0.85) for all samples except S12A and S13A. These results are also

288    found on experimental microstylolites in quartz (Sdiss1 and Sdiss2 in Table 1, Gratier et al.,

289    2005), suggesting that an physical process smoothes the stylolites at small wavelengths.

290    Plotting the results of the analysis in $\mu$ versus H space, one can distinguish between two

291    classes of stylolites at long wavelengths. (Figure 10). A first class, called homogeneous

292    stylolites, contains two kinds of profile: i) the almost-everywhere irregular stylolites

293    (Sjura1 or S12A) and ii) the smooth stylolites (S11C or S10A). For both kinds, the

294    parameter $\mu$ is close to 2 (greater than 1.75), which represents few heterogeneities in the

295    rock. Irregular stylolites have a localized roughness parameter $H$ that varies around 0.5 (0.4

296    to 0.5 in the results obtained here), contrary to smooth stylolites where $H$ is close to 1.

297    Stylolites of this class can be simulated by dynamic surface growth models such as the

298     Langevin growth equations (Renard et al., 2004; Schmittbuhl et al., 2004) because profiles

299     have the same kind of irregularity almost everywhere.

300     The second class of stylolites, called heterogeneous stylolites, contains a variety of

301     morphologies. In this case the parameter $\mu$ is close to 1.5 (stylolites S3b or S0_8). These

302     stylolites are non-stationary. In this case, the initial heterogeneities in the rock that are

303     reached by the stylolite during its propagation are recorded in the stylolitic signal. More

304     exactly, above the millimeter scale, where elastic interactions dominate, heterogeneity may

305     be seen in the signal. Below the millimeter scale, where surface tension dominates, this

306     heterogeneity has disappeared.

307     Agglomerative nesting, clustering methods and principal component analysis (not shown

308     here) have been performed and indeed show that statistical analysis supports the

309     classification of stylolite morphologies in two different classes.

310     *4.3. Simulations*

311     Given a set of parameters ($H$, $\mu$) for both regimes (large and small length scale behaviors

312     from both parts of the cross-over length scale), the behavior of all measured stylolites can

313     be reproduced with two SBPWS. This technique can be used to simulate a wide range of

314     stylolite morphologies (Figures 7, 9):

315     - those which are close to stationary profiles ($\mu$ close to 2);

316     - smooth profiles with $H$ close to 1 to irregular profiles with a fractional exponent $H$;

317     - more heterogeneous profiles with alternating smooth and irregular zones (where $\mu \neq 2$).

318     An interesting perspective would be to use the shape and regularity of stylolites in order to

319     evaluate the heterogeneity of the rock before or during the stylolitic process, and therefore

320     better characterize under which conditions (depth, cohesion of the sediment) stylolites

321     form. Another perspective would be to choose a different noise (a fractional stable noise for

322 instance) in the Langevin growth equations proposed in Renard *et al.* (2004) and

323 Schmittbuhl *et al.* (2004). This remains a real prospect for continuous stylolites models and,

324 more generally, a theoretical extension of rough surface growth models.

## 4. Conclusion

326 When the increments of a mathematical function are not stationary (in other words their

327 statistics vary along the coordinate), or the variance of their distribution is infinite, standard

328 tools (Fourier spectrum or average wavelet coefficient analysis) fail to capture a roughness

329 property from a scaling property.

330 Therefore, an extension of such tools to non stationary signals is proposed here by using a

331 two-parameter approach. One of the parameters, the local roughness exponent $H$, describes

332 the noisiness or waviness of the signal. The second parameter, $\mu$, describes how the

333 statistical properties vary along the signal.

334 Applied to stylolites, two kinds of geometry can be distinguished.

335 i) Stationary stylolites, where the statistics do not vary along the stylolite. For this kind of

336 stylolite, two sub-families can be defined: stylolites that are almost flat everywhere and

337 those that are very wavy everywhere.

338 ii) Non-stationary stylolites where wavy portions alternate with flatter ones. In this case, we

339 propose that heterogeneities initially present in the rock strongly control the stylolite

340 morphology. To our knowledge, this second kind of stylolite, which has fossilized the

341 heterogeneities of the rock in its morphology, has not been previously quantified. Detailed

342 microstructural and chemical mapping studies focusing on the characterization of

343 heterogeneities around stylolites would surely bring new information.

344 This difference between the two families of stylolites is detected only for wavelengths

345 greater than a crossover scale close to the millimeter. Below this scale, the statistics of all

346    the stylolites are very homogeneous, indicating that a physical process, probably driven by

347    the minimization of the local curvature, smoothes the stylolites at small scales.

348

349    **Acknowledgments**

350    This project was supported by the CNRS (ATI and DyETI programs).

351

**Appendix A: Algorithm to build synthetic signals**

```
%//////// Run the styloprocess function //////////////
%Matlab© program to create the stylolites of Figure 7
%Parameters of the simulation
%K: depth of the tree (2^K+1 is the number of points on the
profile)
%mu: heterogeneity parameter (between 1 and 2)
%H: local roughness exponent (between 0 and 1)

function (stylolite) = styloprocess (K,mu,H)

x=linspace(0,2,2^(K-1));
y=1-abs(x-1);
psi=(-y,y,0);
trees=createtree(K,(2-mu));
profile=reconstruct(K,trees,H,psi);
plot(surface);

%//////// Galton-Watson Tree //////////////
function (trees)=createtree(K,p)

randn('state',sum(100*clock));
trees(1)=randn(1);
for m=0:K-1
    for l=0:2^m-1
        zfather=2^m+l;
        zson1=2*zfather;
        zson2=2*zfather+1;
        if (trees(zfather)==0)
            trees(zson1)=0;
            trees(zson2)=0;
        else
            if (rand(1)<p)
                if (rand(1)<1/2)
                    trees(zson1)=randn(1);
                    trees(zson2)=0;
                else
                    trees(zson2)=randn(1);
                    trees(zson1)=0;
                end
            else
                trees(zson1)=randn(1);
                trees(zson2)=randn(1);
            end
        end
    end
end
```

```
400   %/////////////// Reconstruction /////////////////
401   function (sig)=reconstruct(K,trees,H,psi)
402
403   sig=zeros((1, 2^K+1));
404   for m=0:K
405       psim=();
406       for j=1:2^(K-m)+1
407           psim(j)=2^(m/2)*psi(2^(m)*(j-1)+1);
408       end
409       sigtemp=(0);
410       for l=0:2^m-1;
411           zfather=2^m+l;
412           psitemp=2^(-m*(H+1/2))*trees(zfather)*psim;
413           sigtemp=(sigtemp,psitemp(2:2^(K-m)+1));
414       end
415       sig=sig+sigtemp;
416   end
```

## Appendix B: Calculation of $H$ and $\mu$ on 1D signals

A 1-D profile $h(x)$ is observed on a regular grid (at space $x_i = i / 2^n$ for $i = 0 \ldots 2^n - 3$).

Note the second order variation, an approximation of the second order derivative, at point $x_i$, by

$$\Delta_a h(\frac{i}{2^n}) = \sum_{l=0}^{2} a_l h(\frac{i+l}{2^n}) \qquad \text{(B1)}$$

where $a = (a_0, a_1, a_2) = (-1, 2, -1)$.

Summing the $2^n - 3$ variations $\Delta_a h(i / 2^n)$ for $i = 0 \ldots 2^n - 3$, the statistic $V_{n,s}$ is obtained:

$$V_{n,s} = \sum_{i=0}^{2^n-3} \left( \Delta_a h(\frac{i}{2^n}) \right)^s \qquad \text{(B2)}$$

where $s = 2$ (also called quadratic variations) or $s = 4$ (quadric variations). This statistics behave according to a power law depending on the parameters $H$ and $\mu$, with

$$V_{n,s} \approx 2^{n(sH - \log_2 \mu)}.$$

If we note,

429 $$W_{n,s} = log_2\left(\frac{V_{n-1,s}}{V_{n,s}}\right)$$ (B3)

430 then $W_{n,s} \xrightarrow{n \to \infty} sH - log_2\mu$ and by linear combination, either $\mu$ or $H$ is obtained. The

431 estimators are respectively:

432 $$\mu_n = 2^{2W_{n,2} - W_{n,4}} \text{ and } H_n = \frac{1}{2}\left(W_{n,4} - W_{n,2}\right).$$ (B4)

433

493 **Figures & Table**

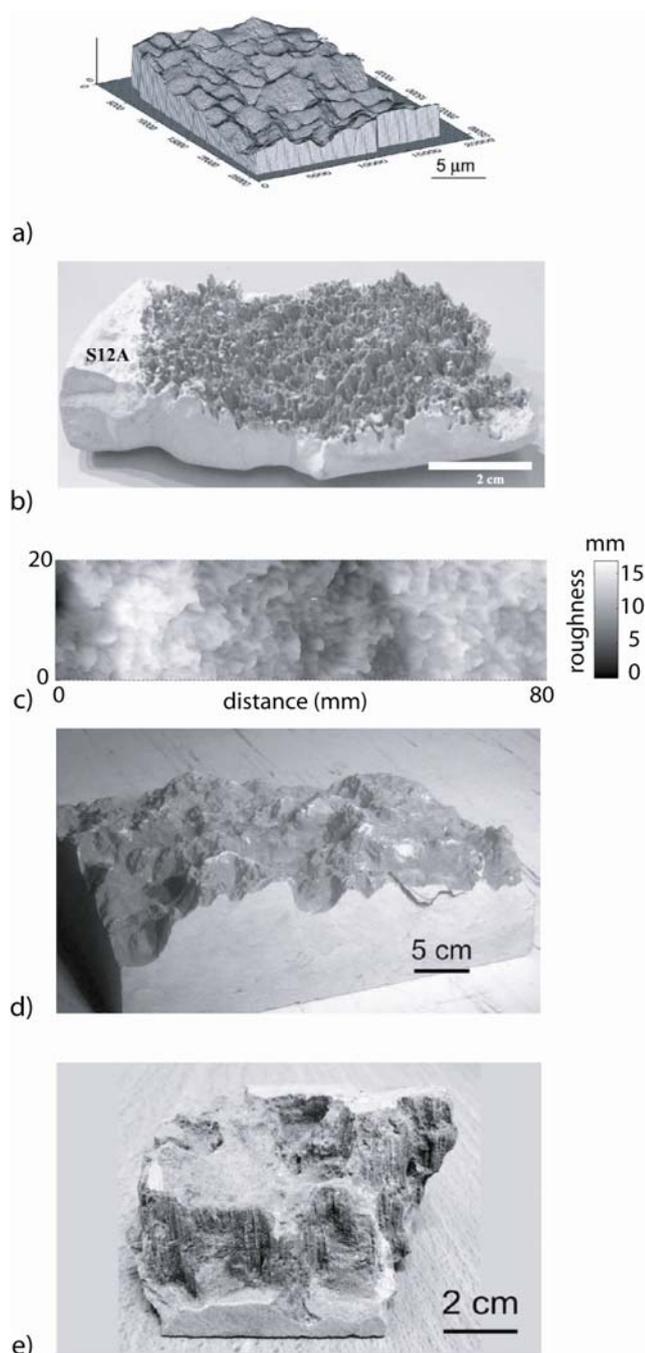

494
495 **Figure 1.** Various shapes of stylolites. a) Digital elevation model of a microstylolites
496 measured at the contact between experimentally deformed quartz grains (after Gratier et al.,
497 2005, isotropic scale). b) 2D stylolite surface S12A in a limestone. c) Roughness field of
498 the surface S12A measured using a laser profilometer (Renard et al., 2004). d) Stylolite S3b
499 showing local variations in roughness, with alternating smooth and rougher areas. Such
500 lateral roughness variations are a good visual indicator that the roughness statistics are not
501 the same all along the profiles. e) Stylolite in limestone with vertical peaks showing strong
502 lateral variations in height. It was not possible to measure the roughness of such stylolites
503 because of local overhangs.

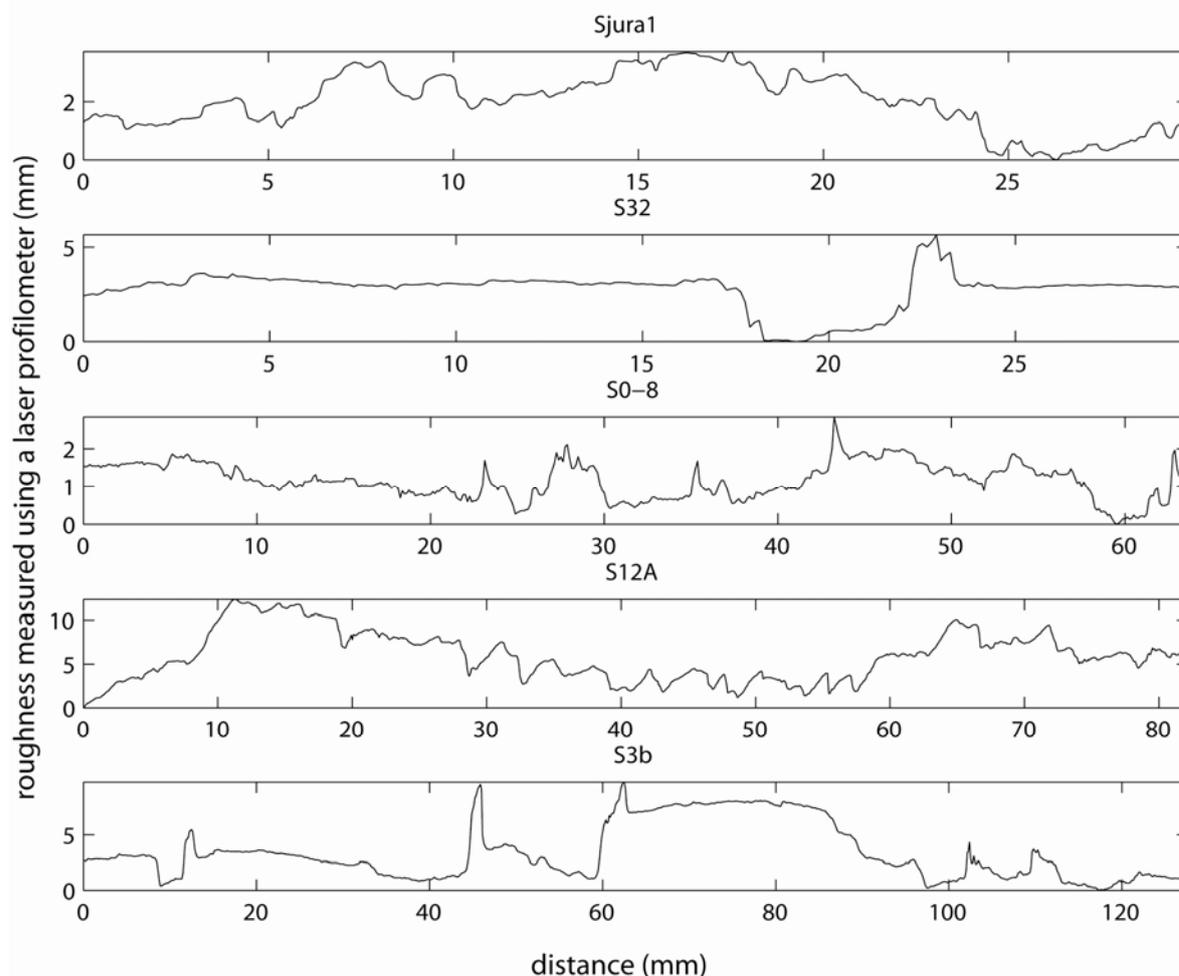

**Figure 2.** Examples of the 1D roughness of different stylolites in limestones measured using laser profilometer (see Renard et al. (2004) for the measurement technique). The waviness of the stylolite, characterized by the Hurst exponent $H$ varies from sample to sample. Moreover, within the same stylolite, regions with smooth or wavy roughness can be defined, and characterized by the amount of irregularities defined by the parameter $\mu$ (see text). Scales are given in mm. The characteristics of each profile are given in Table 1.

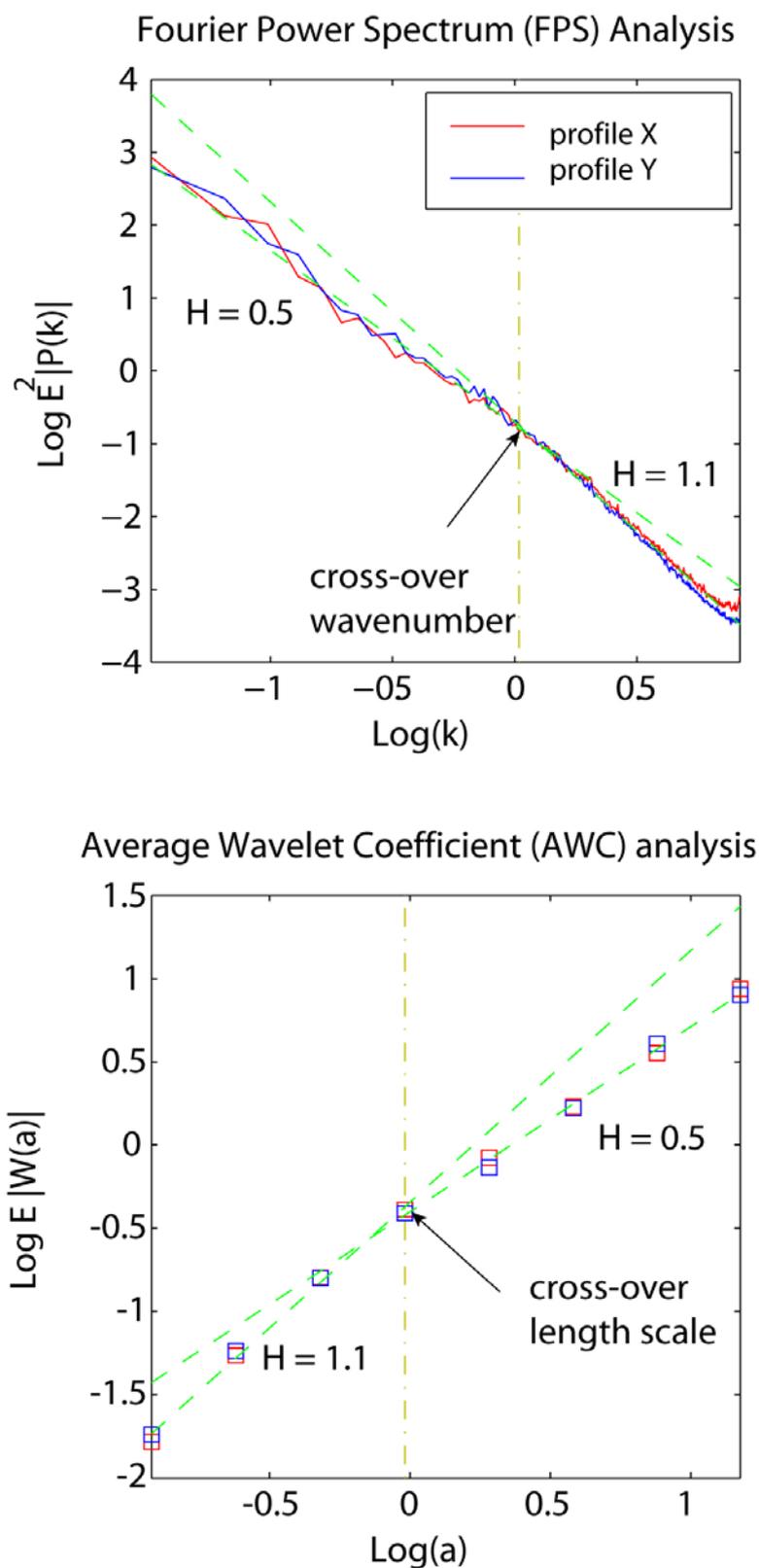

**Figure 3.** FPS (top) and AWC (bottom) for the stylolite Sjura1. These two independent scaling methods show that there is a crossover at ~1mm between the small wavelengths (H~1.1) and the large wavelengths (H~1.5).

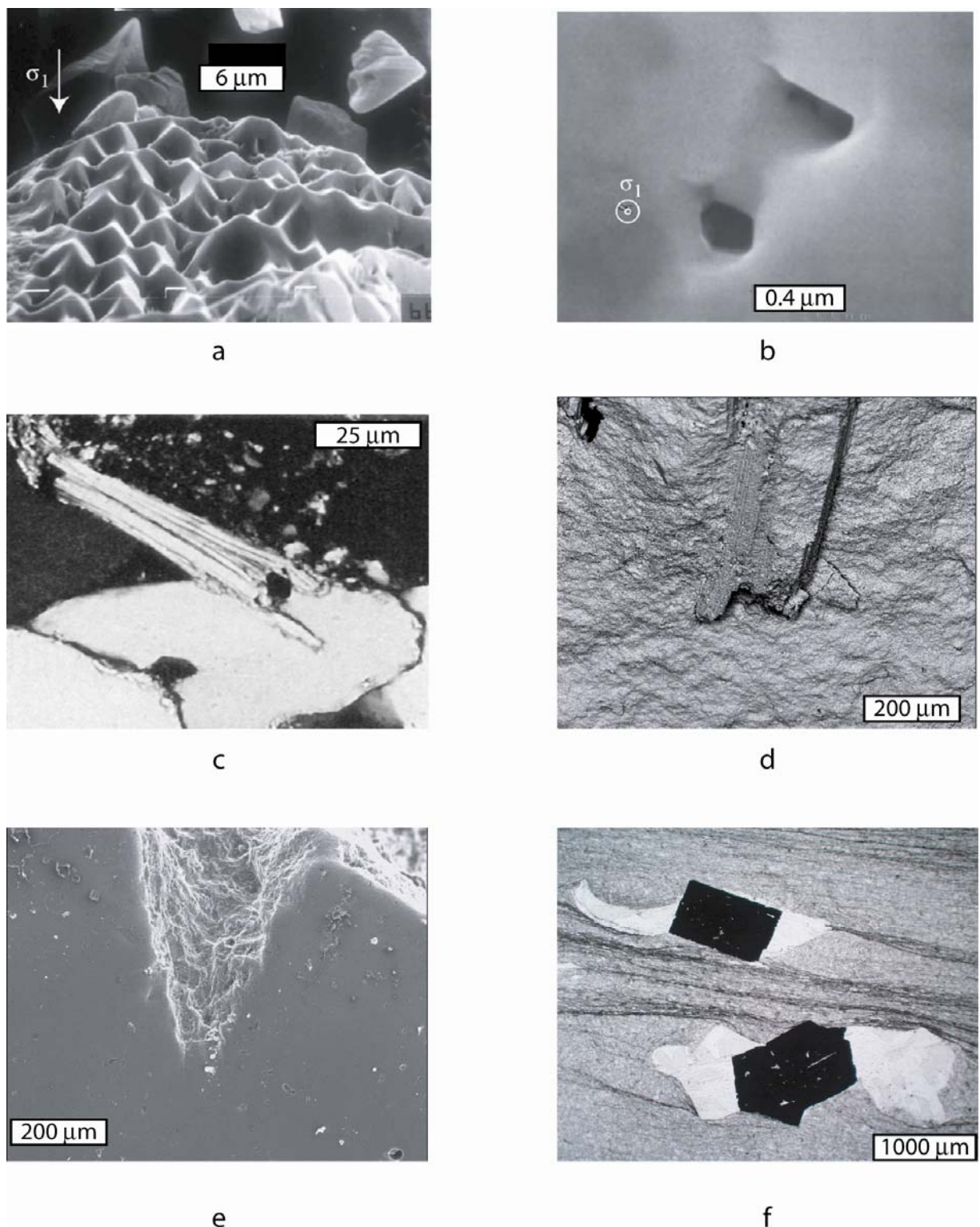

**Figure 4.** Heterogeneities associated with stylolites. a-b) Microstylolite on a quartz grain (Gratier et al., 2005) and zoom on two dislocation pits where deformation is localized. c) Mica indenting a quartz grain in a North Sea Sandstone and showing a wavy interface at the grain scale. d-e) Zoom on stylolite peaks in the sample Sjura1. f) Dissolution seams ("flat" stylolites) deflected by pyrite crystals and quartz pressure shadows in a metamorphic schist from Bourg d'Oisans (Alps, France).

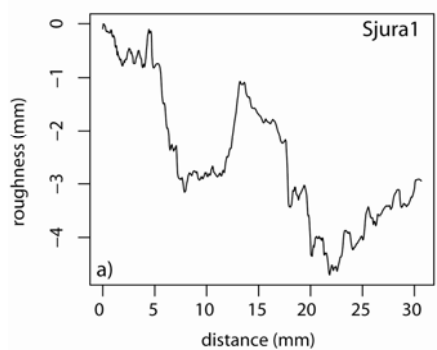

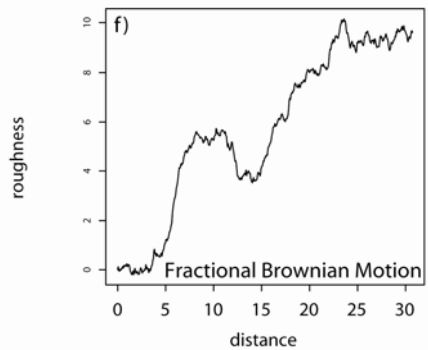

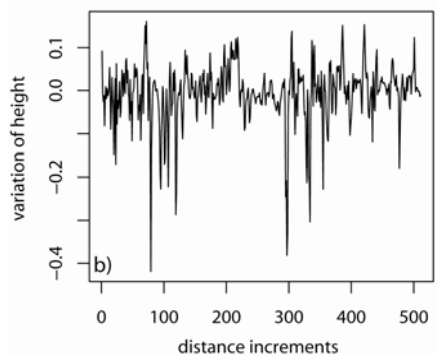

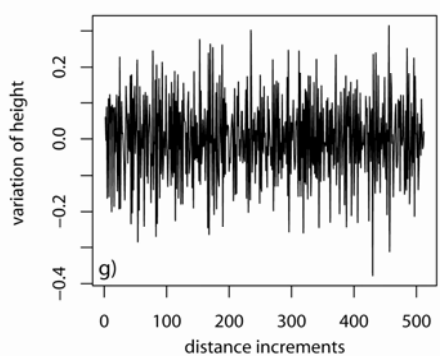

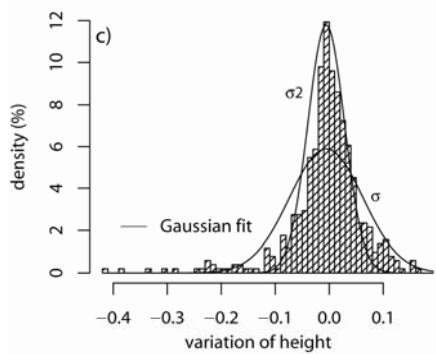

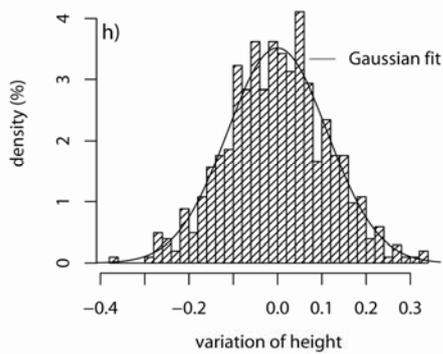

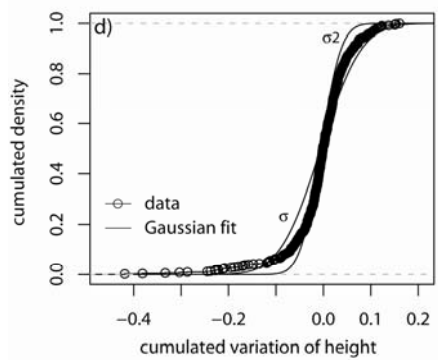

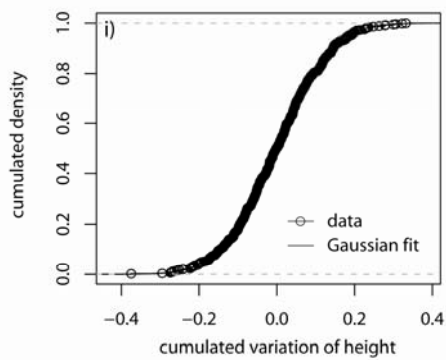

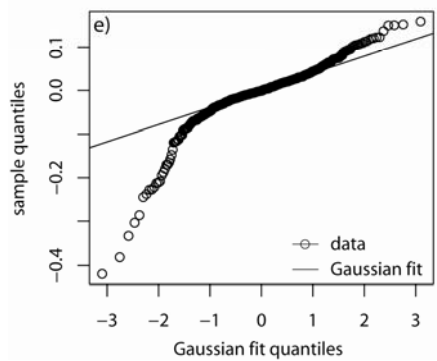

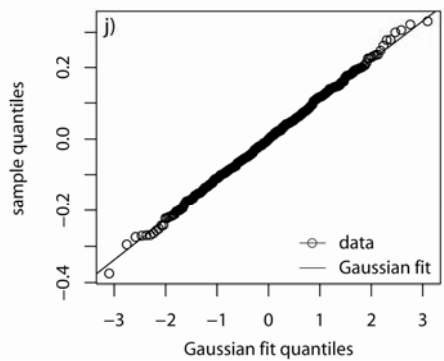

525 **Figure 5.** a) Laser roughness measurement of a 1D profile from the stylolite Sjura1. b)
526 Local increments of the stylolite Sjura1, corresponding to the first order discrete derivative
527 of profile a). c) Histogram of the increments of b) with the best Gaussian fits represented by
528 the two curves, which have the same standard deviation ($\sigma$) and half the standard deviation
529 ($\sigma2$) of the stylolite data. d) Cumulative distribution function of b). The two lines represent
530 the best Gaussian fits as in b). The large jumps of the local increments and the long tails in
531 the histogram cannot be accounted for using Gaussian stationary statistics (plain curves). e)
532 Quantile-quantile plot that adjusts the sample distribution in d) against the best Gaussian
533 distribution. This corresponds to the difference between the data and the Gaussian estimate
534 of d). For a Gaussian distribution a straight line should be observed. f -j) Same plots for a
535 synthetic fractional Brownian motion. In the quantile-quantile plot, the synthetic signal and
536 the Gaussian best fit adjust perfectly on a straight line, showing that the fractional
537 Brownian motion is a Gaussian stationary increments signal.
538

539

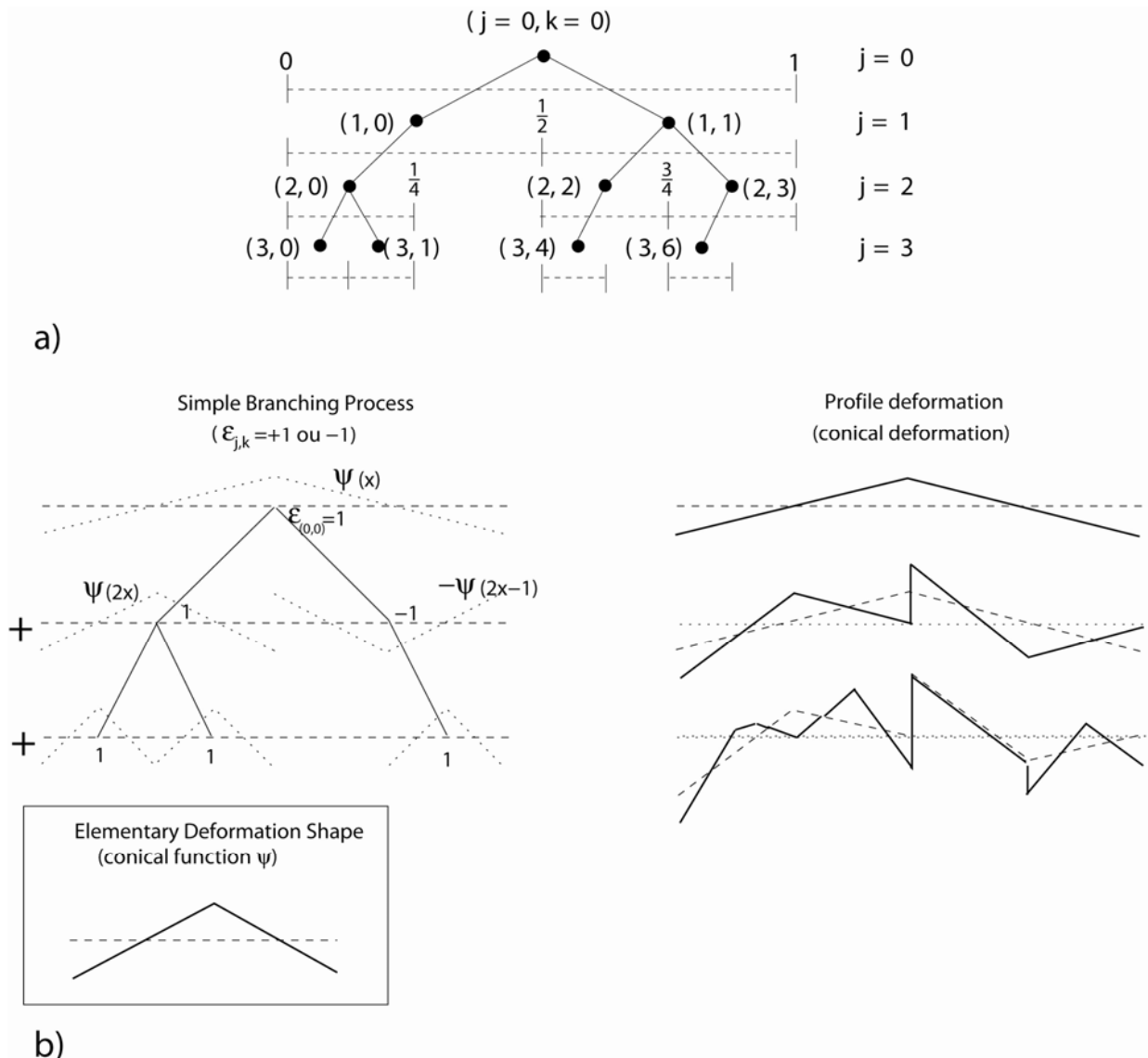

a)

b)

540
541 **Figure 6.** a) Galton-Watson tree (simple branching process) and indexes for the wavelet
542 construction. b) Construction of a synthetic 1D profile using the branching process wavelet
543 series. Such technique is used to build the synthetic signals of Figure 7, using the algorithm
544 given in Appendix A.

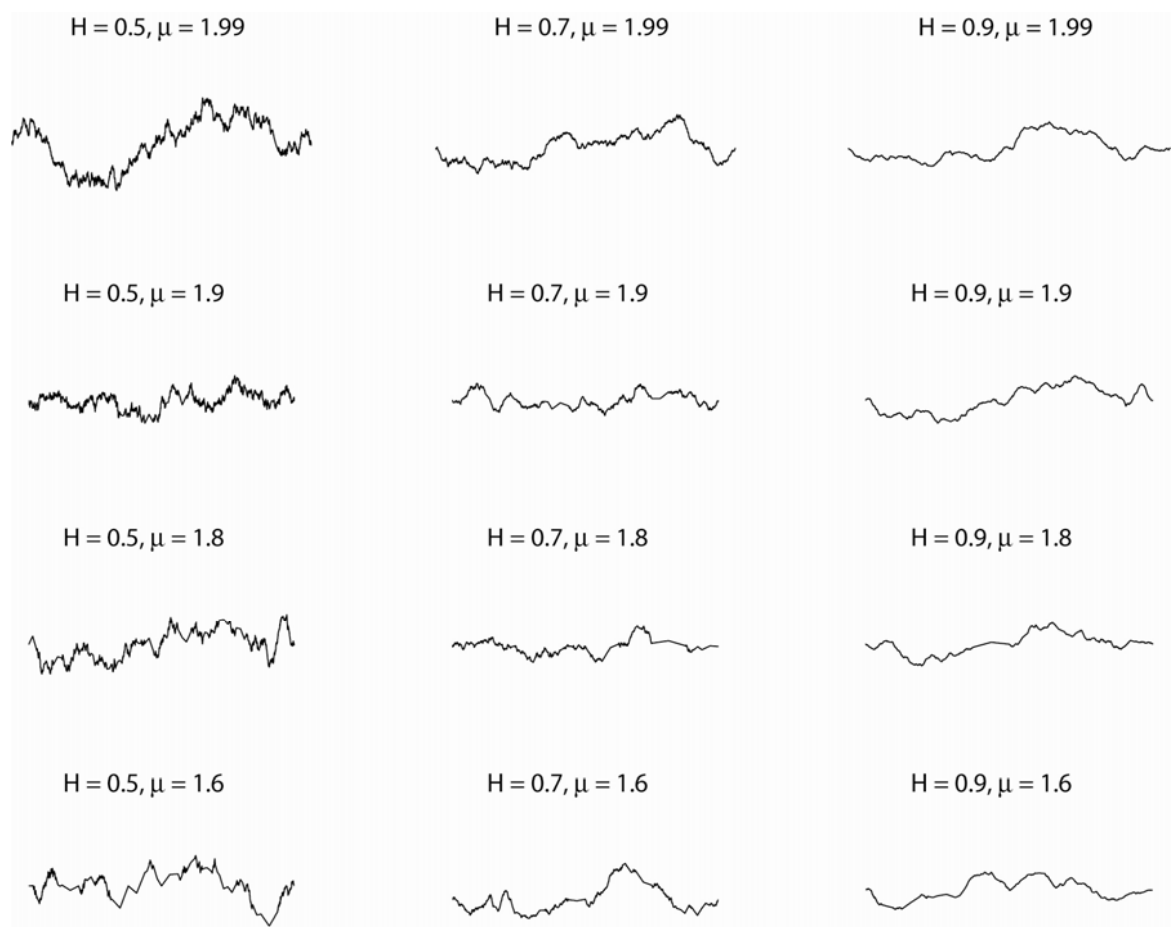

**Figure 7.** Simulated stylolites with statistical roughness properties characterized by two parameters. The variability in the stylolite morphology is controlled by *H* which describes the apparent noisiness (smoothness) of the roughness, and $\mu$ which describes the spatial variability (heterogeneities at all scales) along the stylolite.

552
553

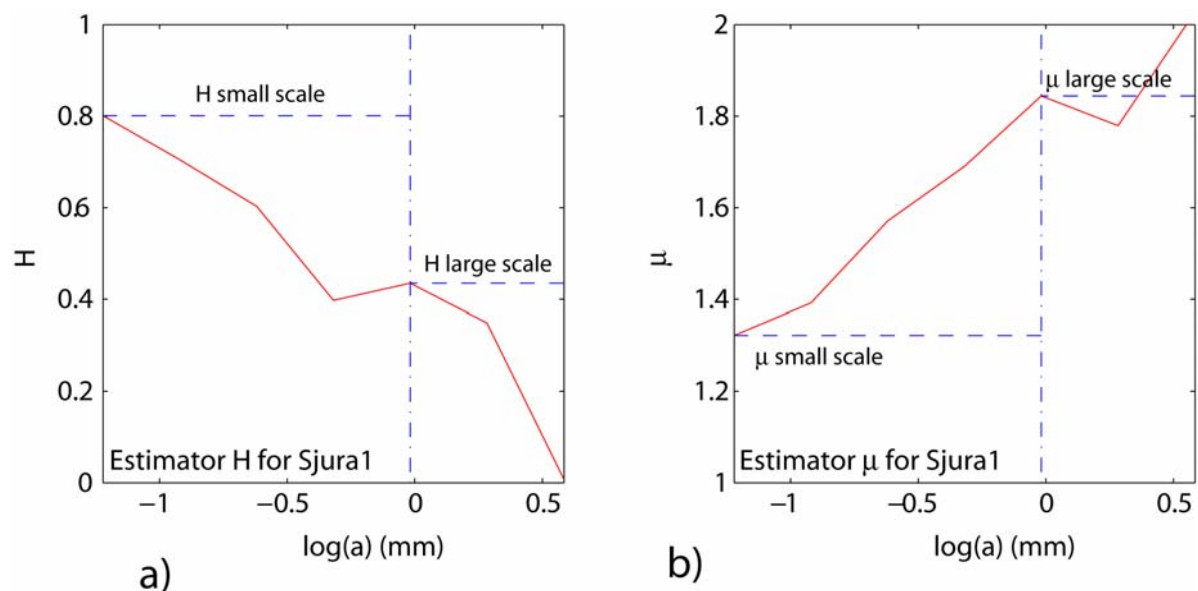

554

a)          b)

555 **Figure 8.** a, b) Estimators of $\mu$ and $H$ for the stylolite Sjura1 at small length scales and
556 large length scale. As the length-scale $a$ decreases ($n$ increases in the equations of Appendix
557 B, where $n$ represents the level of branching in Figure 6), the estimated values converge
558 respectively to $H$ and $\mu$ just above the cross-over length scale for large length scales and as
559 allowed by the precision for small length scales.
560

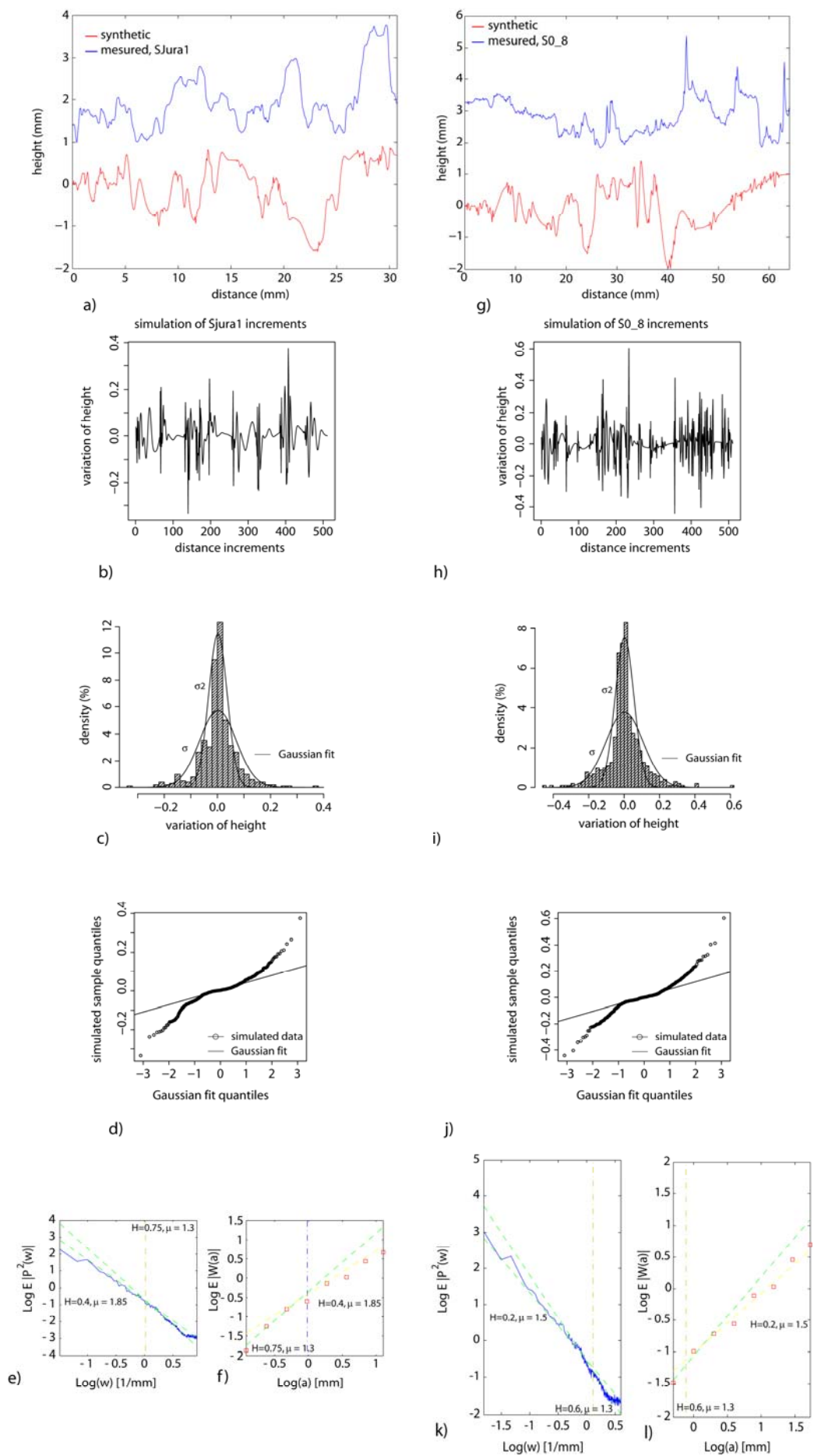

562  **Figure 9.** Using both values $H$ and $\mu$ estimated at large length scale and at small length
563  scale, one can reproduce different morphologies of stylolites using a combination of two
564  SBPWS behaviors. a) Profile of the stylolite Sjura1 (see Table 1) and synthetic profile with
565  the same parameters at small and large length scales as those estimated on Sjura1. b)
566  Derivative of the synthetic signal of a) showing the increments. c) Histogram of the
567  simulated increments. d) Quantile-quantile plot, as in figure 5 showing the departure from a
568  Gaussian distribution. FPS (e) and AWC (f) spectra analysis for the synthetic signal having
569  the same statistical properties as Sjura1. The green dashed straight lines at small and large
570  length scales indicate the estimated slopes, showing the two characteristic slopes and the
571  crossover length scale. g-l) Stylolite S0_8 and synthetic profiles with the same parameters
572  as estimated on S0_8 and similar analysis than in a-d). FPS (g) and AWC (h) spectra of the
573  synthetic stylolite showing the two characteristic slopes and the crossover length scale.
574

575

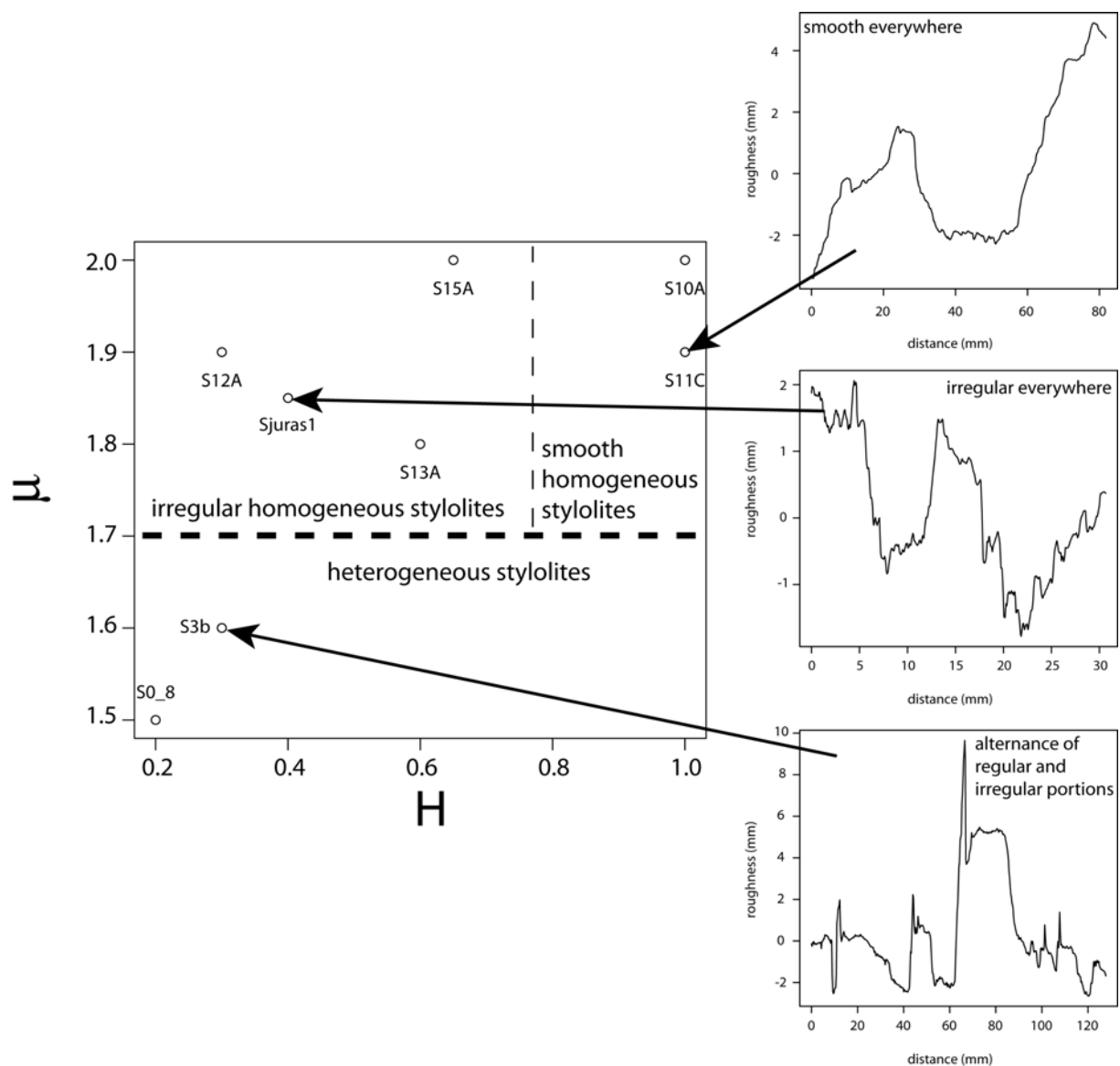

576
577 **Figure 10.** Various morphologies of stylolites based on their statistical properties at large
578 length scale. Two main families can be identified, based on their statistical properties: those
579 which are either regular or irregular everywhere, and those with alternating regular and
580 irregular portions.
581

582    **Table 1.** Large and small length scale scaling exponents of the various stylolites.

| stylolite | origin | $H_{small}$ | $\mu_{small}$ | $H_{large}$ | $\mu_{large}$ |
|---|---|---|---|---|---|
| Sjura1 | Jura mountains | 0.75 | 1.3 | **0.4** | **1.85** |
| S12A | Vercors mountains | 0.2 | 1 | **0.3** | **1.9** |
| S11c | Burgundy mountains | 0.7 | 1.35 | **1** | **1.9** |
| S3b | Chartreuse mountains | 0.5 | 1.4 | **0.3** | **1.6** |
| S15A | Burgundy | 0.6 | 1.4 | **0.65** | **2** |
| S0_8 | Jura mountains | 0.6 | 1.3 | **0.2** | **1.5** |
| S13A | Burgundy | 0.9 | 1.8 | **0.55** | **1.8** |
| S10A | Burgundy | 0.85 | 1.4 | **1** | **2** |
| Sdiss1 | Experimental microstylolite | 0.8 | 1.25 | **-** | **-** |
| Sdiss2 | Experimental microstylolite | 0.75 | 1.2 | **-** | **-** |

583    1 For more details on the geological characteristics and composition of the stylolites, see Renard et al. (2004)

584    and Gratier et al. (2005) for the experimental microstylolites.